\documentclass[sigconf]{acmart}

\AtBeginDocument{%
  \providecommand\BibTeX{{%
    \normalfont B\kern-0.5em{\scshape i\kern-0.25em b}\kern-0.8em\TeX}}}

\setcopyright{acmcopyright}
\copyrightyear{2018}
\acmYear{2018}
\acmDOI{XXXXXXX.XXXXXXX}

\acmConference[Conference acronym 'XX]{Make sure to enter the correct
  conference title from your rights confirmation emai}{June 03--05,
  2018}{Woodstock, NY}

\acmPrice{15.00}
\acmISBN{978-1-4503-XXXX-X/18/06}

\begin{document}

\title{Integrating Voice-Based Machine Learning Technology into Complex Home Environments}

\author{Ye Gao}
\affiliation{%
  \institution{University of Virginia}
  \country{United States}}
\email{yg9ca@virginia.edu}

\author{Jason Jabbour}
\affiliation{%
  \institution{Harvard University}
  \country{United States}}
\email{jasonjabbour@g.harvard.edu}

\author{Eunjung Ko}
\affiliation{%
  \institution{Ohio State University}
  \country{United States}}
\email{ko.363@buckeyemail.osu.edu}

\author{Lahiru Nuwan Wijayasingha}
\affiliation{%
  \institution{University of Virginia}
  \country{United States}}
\email{lnw8px@virginia.edu}

\author{Sooyoung Kim}
\affiliation{%
  \institution{Ohio State University}
  \country{United States}}
\email{kim.8201@buckeyemail.osu.edu}

\author{Zetao Wang}
\affiliation{%
  \institution{New York University}
  \country{United States}}
\email{zw3478@nyu.edu}

\author{Meiyi Ma}
\affiliation{%
  \institution{Vanderbilt University}
  \country{United States}}
\email{meiyi.ma@vanderbilt.edu}

\author{Karen Rose}
\affiliation{%
  \institution{Ohio State University}
  \country{United States}}
\email{rose.1482@osu.edu}

\author{Kristina Gordon}
\affiliation{%
  \institution{University of Tennessee}
  \country{United States}}
\email{kgordon1@utk.edu}

\author{Hongning Wang}
\affiliation{%
  \institution{University of Virginia}
  \country{United States}}
\email{hw5x@virginia.edu}

\author{John A. Stankovic}
\affiliation{%
  \institution{University of Virginia}
  \country{United States}}
\email{jas9f@virginia.edu}


\begin{abstract}
To demonstrate the value of machine learning based smart health technologies, researchers have to deploy their solutions into complex real-world environments with real participants. This gives rise to many, oftentimes unexpected, challenges for creating technology in a lab environment that will work when deployed in real home environments. In other words, like more mature disciplines, we need solutions for what can be done at development time to increase success at deployment time. To illustrate an approach and solutions, we use an example of an ongoing project that is a pipeline of voice based machine learning solutions that detects the anger and verbal conflicts of the participants. For anonymity, we call it the XYZ system. XYZ is a smart health technology because by notifying the participants of their anger, it encourages the participants to better manage their emotions.  This is important because being able to recognize one's emotions is the first step to better managing one's anger. XYZ was deployed in 6 homes for 4 months each and monitors the emotion of the caregiver of a dementia patient. In this paper we demonstrate some of the necessary steps to be accomplished during the development stage to increase deployment time success, and show where continued work is still necessary. Note that the complex environments arise both from the physical world and from complex human behavior.


\end{abstract}




\maketitle

\section{Introduction}
\label{sec:introduction}
Smart health research teams often develop novel machine learning technologies. To prove the value of the technologies, the research teams have to deploy the solutions into a complex environment such as a smart home \cite{bedon2020home, costin2009telemon, khan2017smart, gao2020monitoring}. However, the aforementioned works do not describe the problems related to the transition from the development stage to the deployment stage. In other words, during the development stage, the research teams try to develop their solution in an environment, usually a lab environment and/or a controlled home environment, which is less complex than the environment in which the technology is going to be deployed. It is well known that when new technology is actually deployed in real complex environments, issues previously unseen in the less complex environments  are going to occur, especially for long-term deployments. This paper describes the XYZ System, a pipeline of voice-based machine learning solutions for in-home monitoring of the emotion and verbal conflict experienced by a caregiver of a dementia patient. XYZ is considered a smart home technology because it helps caregivers of dementia patients better manage their anger by notifying them when an anger or verbal conflict is detected. As Pena et al. \cite{pena2015integrating} have discussed, one school of promoting emotion regulation is to permit the individuals from experiencing anger. Since people can be bad at recognizing their own emotions \cite{goerlich2018multifaceted}, a smart health technology that notifies them when an anger or verbal conflict arises is very important to their emotion regulation.

The goals of this paper are to (1) demonstrate solutions for what are needed at pre-deployment time to increase the success of voice-based classifiers once deployed, and (2) present lessons learned for environmental and behavioral complexities uncovered \textbf{during 4 month deployments in six homes}. 
The main challenges of integrating novel machine learning technology into complex environments are that both the physical environments and the participants' behaviors are complicated. For a smart health technology that uses voice for emotion detection \cite{gao2021emotion}, the environment is complicated because the solution needs to address the problems of acoustic signal's deterioration due to background noise such as birds chirping, room reverberation due to the signal bouncing off the surface of furniture, and deamplification as a result of the distance between the human speaker and the microphone. We call these environmental distortions acoustical realisms. Also, the television could be on and the voice from the television adds to the complexity of detecting the participants' emotions. Human behavior is very complicated, too. For example, people could be yelling across the room to ask each other what is for dinner. They are not necessarily angry, but yelling itself is seen as an anger event by typical emotion detection technology. Visitors are common and for privacy the caregiver may turn the system off and forget to turn it back on. 

The gist of our solutions at pre-deployment time to ensure maximizing the post-deployment success of the machine learning algorithms in the XYZ System is that we make sure to evaluate the algorithms on data samples that are most similar to the samples that the algorithms are about to encounter in the designated environment in which the algorithms are/will be deployed. Using the XYZ System as an example, the (acoustical) samples that the algorithms in the XYZ System are about to encounter in its designated environment (a participant's home) are voice samples that are environmentally distorted. It is worth-noting that some works, such as Chen et al. \cite{chen2019arasid}, do consider the designated environment in which their algorithm is going to be deployed during the pre-deployment stage. However, they do not evaluate if their hypothesis that the algorithm \emph{indeed performs well during post-deployment time} holds true. Unlike works such as Chen et al. \cite{chen2019arasid}, we perform evaluations to show that after our rigorous pre-deployment time assessment of the algorithms to be deployed in the designated environment, our chosen algorithms indeed perform well during the post-deployment stage when they are being deployed in the designated environment (the home of a patient-caregiver dyad), which is one novelty of this paper. A more nuanced finding of this paper is that different acoustic components need to address different environmental realisms. For example, the voice activity detection (VAD) model only needs to deal with non-speech background noise, reverberation, and deamplification, whereas the emotion detection model needs to also deal with background speech (such as those from the tv) because background speech could confound the classification of the emotion detection model which uses speech to determine the emotion in that speech.

The purpose of the XYZ System is that we want to help caregivers of dementia patients to manage their anger as well as the verbal conflict events that arise during care-giving. To achieve these goals, the XYZ System has three add-on components: the Cloud to which the audio clips are uploaded for safe keeping, the M2G monitoring component that notifies the developer(s) if the XYZ System crashes so the developer(s) can investigate the causes, and the Recommender System using EMA, which sends messages that are interventions or recommendations to help the caregivers manage their emotions and verbal conflict, if anger or verbal conflict are detected by the XYZ System. The entirety of the XYZ System and the three add-on components are referred to as the \emph{XYZ-W} (XYZ-Whole) System.  We also present some lessons learned on several aspects of \emph{XYZ-W}. 

Note that the focus of this paper is the XYZ System, not its add-on components, so we do not evaluate how effective the add-on components are at helping the caregiver manage their emotions and verbal conflicts that arise during care-giving. Nonetheless, given their essential role at helping the XYZ System achieve its goals, we have decided to include some discussion of the add-on components in the paper.

The contributions of this paper are based on the in-home deployed XYZ-W system from \textbf{six completed 4-month deployments of real caregiver-Alzheimer's patient interactions}. This work was performed under an approved IRB. The main contributions are:

\begin{itemize}
    \item A pre-deployment approach for improving standard and novel acoustic technologies deployment time success. 
    
    \item A pre-deployment approach to make components in the deployed technology adaptable to they can be easily adjusted during the deployment time to satisfy the needs of the participants.

    \item Developing a set of privacy mechanisms that are shown to help dyad recruitment and satisfy privacy concerns of  actual users of the system. 
    
    \item Our post-deployment validation indicates that people are not always good at recognizing their own emotions, which reconfirms similar findings in the literature \cite{goerlich2018multifaceted}.

\end{itemize}
\section{Related Works}
\label{sec:related_works}
In this related works section, we discuss existing smart home technologies as well as how human behavior affects the efficacy of smart technologies. We also briefly talk about emotion detection and conflict detection using voice, since the XYZ system uses the voice of its participants to detect if they are angry and if they are in a verbal conflict.

\subsection{Methodology for Deploying Smart Health Technologies}

Zeadally et al. \cite{zeadally2019smart} argue that deploying smart health technologies can improve healthcare availability, cost, and access. However, there have not been enough works for the methodology for deploying smart health technologies (usually in smart homes). Bellandi et al. \cite{bellandi2021design} is among the first works that propose a design methodology that matches smart health requirements. They argued that due to the rapidly aging world population, smart health technologies are about to playing an even more important role in caring for the aging population at homes. They propose that the perspective of deploying those smart health technologies at homes should be user-centered. To select a solution to be deployed, researchers need to perform domain analysis (surveying the state of the art), requirement gathering (stakeholder elicitation), requirement integration (determining the consistency of the requirements), and then finally solution selection. 


\subsection{Human behavior that affects the efficacy of smart technologies}
There has not been enough work on how human behavior affects the efficacy of smart technologies. A previous work \cite{yaghoubi2010factors} hypothesizes that the perceived usefulness directly contributes to the user's attitude towards the technology. Nikou et al. \cite{nikou2019factors} suggest that perceived innovation and cost directly contribute to if the user will adopt a smart technology. Perri et al. \cite{perri2020smart} point out that the attitude of the user directly affects the adoption of smart technologies, which we would like to extend to that their attitude directly affects the efficacy of smart technologies as shown by the observation of a user of our XYZ system: he was not very compliant and sometimes pulled out the wires to the laptop on which we ran our system. As a result, the computer was out-of-power and the system was unable to monitor the participant's emotion for a period.

\subsection{Voice Activity Detection}
There have been lots of work on voice activity detection (VAD) so we only discuss the recent advances in the field. MarbleNet \cite{jia2021marblenet} uses deep residual network consisting of blocks of 1-D time-channel separable convolution, able to achieve the state-of-the-art performance with the advantage that the number of their parameters is significantly smaller. The robustness of MarbleNet is also extensively studied to demonstrate that it is robust to real-world acoustical distortions. Using teacher-student training, Dinkel et al. \cite{dinkel2021voice} also strive to train a model that is robust to real-world acoustic distortions. Dinkel et al. identify that traditional VAD algorithms are trained on data devoid of such acoustic distortions, and therefore their usage is limited to data without the acoustic distortions that are inevitable in the real world, rending them unable to perform well in real-world settings. Other works on VAD include Wang et al. \cite{wang2022cross} that uses a cross channel attention based model to achieve voice activity detection in the M2met challenge, Braun et al. \cite{braun2021training} that is specifically concerned about dealing with the robustness issue of many state-of-the-art models. What is worth-noting is that, some works developed for other purposes such as transcription, can be used as voice activity detection models. For example, the Google speech Recognition (GSR), a transcription service, outputs the transcribed sentence from an audio clip if that audio clip is speech, and it will throw an exception is the audio clip is silence. It is worth noting that although works such as MarbleNet \cite{jia2021marblenet} and Dinkel et al. \cite{dinkel2021voice} attempt to ensure that they work on datasets that account for realism to be encountered in real, designated environments in which the algorithms are to be deployed, they do not evaluate their post-deployment performances in the real, designated environments. Realisms that the VAD model deals with usually arise from background noise such as footsteps, and the VAD model needs to differentiate not only silence from human speech but also those background noises from speech. The realisms that the VAD model faces is simpler than the models that we discuss in the later sections, the speaker identification (SID) model, the emotion detection model, and the conflict detection model, which needs to deal with the tv sound as the speech from the tv could affect the classification performance of these models.

\subsection{Speaker Identification}
Again, the works in the field of speaker identification (SID) are abundant, so we only discuss the recent advances in the field. Chen et al. \cite{chen2021graph} introduce a graph-based speaker identification model that is reliant on speaker label inference. It is particularly concerned with the task of SID in household scenarios. WavLM \cite{chen2022wavlm} recognizes that the speech content by by speakers contains multi-faceted information such as the identities of the speakers, the content of the speech, and paralinguistics. WavLM is propsoed as a pre-trained model that can be used to be fine-tuned for the purpose of various speech recognition tasks such as speaker identification. Snyder et al. \cite{snyder2018x} proposes an xvector, the results of mapping variable-length spoken clips to fixed-dimensional embeddings. Again, works such as Chen et al. \cite{chen2021graph} evaluate their algorithms on datasets in which the realism to be encountered in real, designated environments in which the algorithms are to be deployed, but no post-deployment evaluation is presented in such works to show if their approaches to deal with the realism are successful. Realisms that the SID model faces arise from background noise, especially the tv sounds. Note that the realisms that the SID model needs to deal with are more complex than the realisms that the VAD model needs to deal with, as voice from the tv could confound the model from correctly identifying the identity of the speaker in an audio clip. In other words, the SID model needs to be able to deal with more complex background noise (more complex acoustical realisms) than the VAD model.

\subsection{Emotion Detection}
There have also been a lot of work \cite{ danisman2008emotion, emodbdeng2017universum, emodbwang2015speech, emodbalex2018utterance} on using voice as a modality to classify emotions. These use several datasets that have speech files with emotion labels. EMO-DB, a dataset of the German language \cite{burkhardt2005database}, is a popular one for many works \cite{emodbtriantafyllopoulos2019towards, dickerson2014resonate}. It consists of six emotion categories such as anger, sadness, and happiness performed by actors. Another popular dataset is RAVDESS \cite{livingstone2018ryerson} that consists not only of the emotional utterances, but also video footage of emotional speech. CREMA-D \cite{cao2014crema}, like RAVDESS, contains both audio and audio-video samples. Despite the variety in modalities, since some datasets like EMO-DB have only audio samples, and others like RAVDESS have both audio and video samples, the emotions that they have are largely in common. The common emotion categories in the datasets are happiness, anger, sadness, and neutrality. One common problem with the emotional utterance datasets is that they are often collected in controlled studio environments in which realisms expected in a real, designated environment do not exist. Algorithms trained on those samples are shown not to be robust when deployed in the real world \cite{gao2021emotion}. There have been works that attempt to address environmental distortions such as reverberation, deamplification, and the background noise at pre-deployment time \cite{salekin2017distant, wijayasingha2021robustness, shchetinin2020deep, gao2021emotion}. However, they do not confirm at post-deployment stage if their strategies of addressing the realisms work. The realisms that the emotion detection model needs to address, in addition to deamplification, reverberation, and common indoor background noise, are tv sounds. An audio clip produced by a registered speaker could include background sounds that are from the tv, and the speech from the tv could confound the emotion detection model from making the correct prediction as the speech from the tv could be of a different emotion than the speech produced by the registered speakers.

\subsection{Conflict Detection}
There have been several attempts to detect verbal conflict using sound signals that a microphone picks up from the ambient environment. A work \cite{lefter2017aggression} creates verbal conflict between pairs of a student and an actor who act out conflict. From the generated conflict episode, it is observed that overlapped speech is an important indicator of interpersonal conflict \cite{lefter2017aggression}. However, they did not create a model of automatic conflict detection based on their conclusion. Based on the fact that repetition of parts of speech, such as syllables, phrases and words, is indicative of interpersonal conflict, another work \cite{letcher2018automatic} develops a repetition detection model that uses the audio files collected by the on-body sensors of police officers to detect conflict. However, the interpersonal conflicts that police officers encounter during their jobs are not the same as every-day interpersonal conflicts that take place in households between arguing family members. The state-of-the-art modules on automatic conflict detection using speech \cite{caraty2015detecting, grezes2013let}, achieve satisfactory performance on their respective datasets, but their approaches are not evaluated to demonstrate if acoustic distortions of noise, distance, and reverberation affect the results. As a result, the automatic detection of every-day harmful interpersonal conflicts among people in home environments remains unsolved. Again, in addition to the realisms such as reverberation, common indoor background noise, and deamplification, the conflict detection model, just as the emotion detection model, needs to deal with the realisms that are the tv sound: the characters on the tv might be in a verbal conflict (as the background sound for the participants whose conflict we want to monitor), which can confound the conflict detection model.
  
\section{An Overview of the XYZ System}
\label{sec:acoustics}
\begin{figure}
    \centering
    \includegraphics[width=0.5\textwidth]{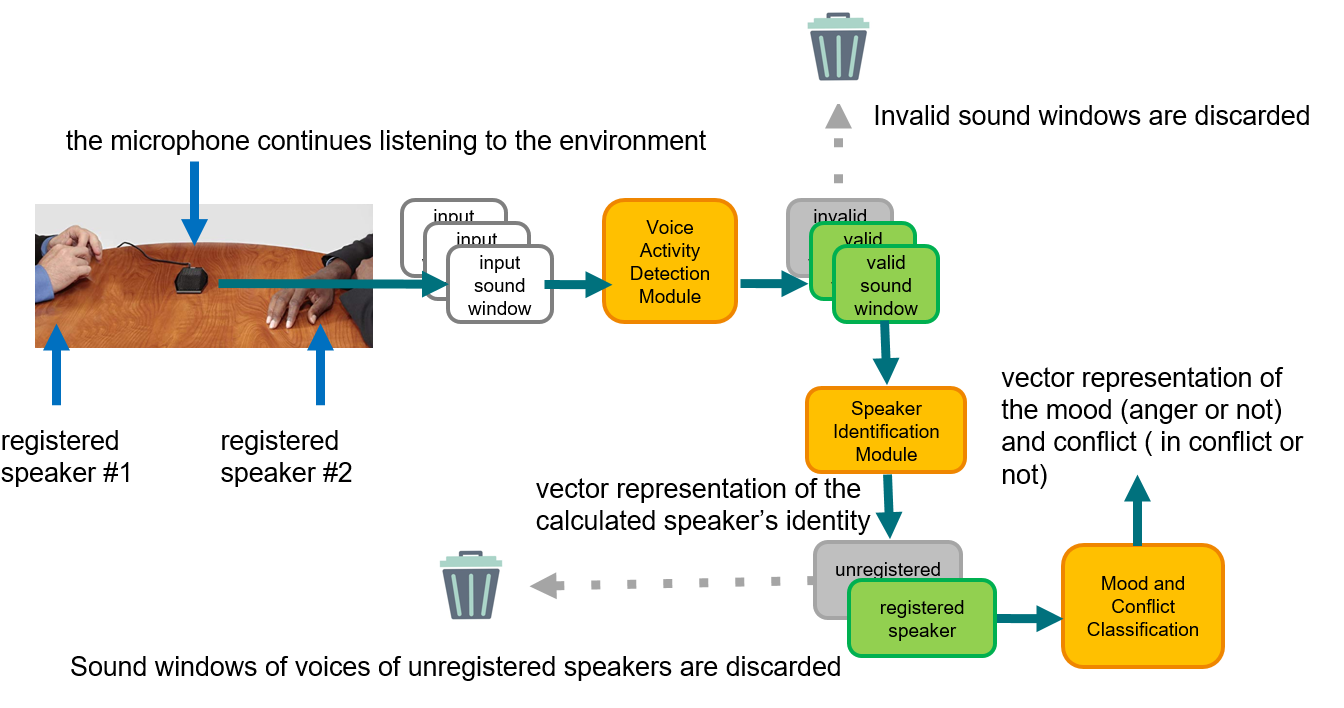}
    \caption{Overview of the XYZ system comprised of four main components: the voice activity detection (VAD) model, the speaker identification (SID) model, the emotion detection model, and the conflict detection model.}
    \label{fig:XYZ_acoustic}
\end{figure}

In this section, we describe the XYZ System, which includes the voice activity detection (VAD) algorithm, the speaker identification (SID) algorithm, the emotion detection algorithm, and the conflict detection algorithm. Three out of the four algorithms are off-the-shelf. The VAD algorithm is Google's transcription services. The SID algorithm is the WavLM algorithm developed by Microsoft \cite{chen2022wavlm}. The emotion detection algorithm is developed by SpeechBrain \cite{ravanelli2021speechbrain}. We developed the conflict detection by ourselves, and the reference to the conflict detection paper is left blank to preserve our anonymity. 
In later sections of this paper, we thoroughly tested the off-the-shelf algorithms to make sure that they meet our needs in the XYZ system. We also argue in later sections that if there exist algorithms that can satisfy the needs of a system, there is no need to develop those algorithms again by the researchers themselves; time and resources can be spared to develop the algorithms that are needed by the system but not available as off-the-shelf ones (such as our conflict detection model). In other words, there is no need to reinvent the wheel. 

In Figure \ref{fig:XYZ_acoustic}, the microphone placed in a central place in a room constantly listens to the ambient environment. The audio stream is sliced into 5-second audio clips and send to the voice activity detection (VAD) model to decide if a given clip contains human voice. The choice of each audio clip being 5-second is based on the observation from previous works that 5-second is long enough to be indicative of the speakers' emotions \cite{gao2021emotion}. The XYZ System discards those audio clips that are invalid; i.e., they contain no human voice. The valid audio clips are sent to the speaker identification (SID) model to detect if the audio clips contain the voice of registered speakers. If yes, they are sent to the emotion detection and conflict detection models, in parallel. The two models each produce vectors in respect to if the speaker(s) in an audio clip is/are angry or is/are in a verbal conflict.

Note that the purpose of this paper is to demonstrate the necessary steps to be accomplished during the development stage before deployment to increase deployment time success, and show where continued work is still necessary. The purpose of this paper is not about how novel the algorithms used in the XYZ System are, although we show that the algorithms are effective at doing their respective jobs and satisfying the research needs.


\section{The Add-on Components to the XYZ System}
\label{sec:XYZ-W_system}
\begin{figure}
    \centering
    \includegraphics[width=0.5\textwidth]{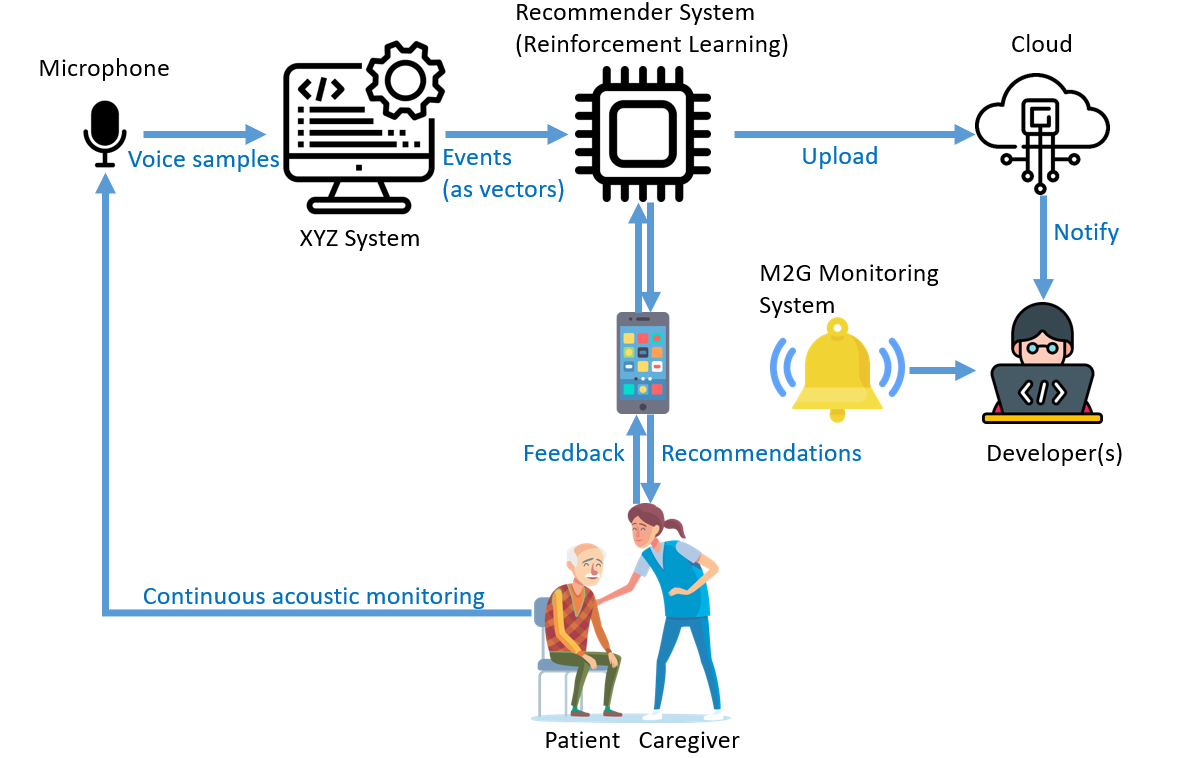}
    \caption{Overview of the XYZ system with \emph{add-on components}. The add-on components are the Recommender System, the Cloud, and the M2G (Monitor) system.}
    \label{fig:XYZ_w_overview}
\end{figure}
In this section, we describe three add-on components for the core XYZ system. Note that the efficacy and effectiveness of the add-on components are not included in this paper because the main focus of the paper is the XYZ system. The add-on components exist to assist the XYZ system by storing the audio clips classified by the XYZ system (in the cloud), sending the identified anger to the caregiver for them for them to be more attentive to their own emotions, and notifying the developers when the XYZ system crashes. In addition to those assistive functionalities, one of the add-ons, the Recommender System, recommends tips to the caregivers to help them manage their anger. The entirety of the XYZ system and its add-ons is referred to as the \emph{XYZ-W} (XYZ-Whole) System. An overview of the XYZ-W system is described in Figure \ref{fig:XYZ_w_overview}.

One of the add-on components is the Recommender System with the EMA. Recommender system uses reinforcement learning to decide if a conflict/anger management tip should be sent and what tip to send. The EMA is a smartphone app developed to send messages to the caregivers. Another add-on component, the Cloud keeps track of the raw acoustic signal files, the SID module's detection on the identity of the speaker, the emotion classification module's classification result, the conflict detection module's detection result, and the recommendations and user responses. 

The final add-on component is the M2G System \cite{ma2017m} that monitors the other three components to make sure that they are alive. If one or more of the other three components crash, the M2G system restarts the component(s) and sends the developers an email to announce the crash so the developers can investigate the causes.

\section{Evaluations}
\label{sec:evaluation}
Many speech processing works have collected or augmented datasets with real world sounds. Good solutions are usually then developed. But, in many cases the resultant datasets have limited in-the-wild sounds so where they actually work is limited, and many times the solutions are not validated in-the-wild, but only on the datasets. 

Our pre-deployment strategies are to significantly increase the collected and augmented datasets to increase comprehensiveness of the situations which are modeled, and specialize the datasets to a specific speech module (if needed).  In pre-deployment, we also choose a set of “best” solutions from the literature and using the comprehensive datasets, determine which ones work. If one or more off-the-shelf solution work, the best one is chosen. Below we show that often many off-the-shelf solutions do not work, but for many common speech tasks there are good solutions available, but it is necessary to verify that. We don’t want to reinvent the wheel. Then, in post deployment we validate the resulting performance on real data, a step often not performed.

Speech processing is typically performed in a pipeline depending on the overall purpose. In our system we need voice activity detection, speaker identification,  emotion detection, and conflict detection. There are commonalities and unique aspects to each of these stages so the pre-deployments strategies must account for them. We now consider each of these 4 stages in more detail. Specifically, we briefly reiterate what the components are, why they are important, and demonstrate what we did at pre-deployment time to maximize the deployment time success based on post-deployment data (how we evaluated them at pre-deployment time). In all cases the acoustic solutions had to address the real world complexities such as reverberation, deamplification, and the many types of noise found in real homes over long deployments. The relative success of the acoustic solutions (at the post-deployment time) confirms our hypothesis that it is possible to perform comprehensive and realistic pre-deployment testing to increase post deployment success.

\subsection{Voice Activity Detection}
The VAD model filters silence and other sounds that are not produced by the human vocal tract. Since the acoustic system is constantly listening to the environment, we do not want to activate the emotion and conflict classifiers when an input sound window contains no human speech. As a result, the VAD model is an important and necessary component in the XYZ System. In this Section, we describe testing on the VAD model for the question: Is it possible to perform comprehensive and realistic pre-deployment testing to improve post-deployment success? Note that the acoustic realisms that the VAD model faces are deamplification, reverberation, and (non-speech) background noise, so in our pre-deployment stage assessment we seek to find a VAD solution that is robust against these three types of acoustical realisms.

\subsubsection{Pre-deployment Stage Assessment}

During pre-deployment time, we first looked into several state-of-the-art VAD algorithms. In particular, we studied the performance of a set of SOTA VAD algorithms on the Aurora-2 database \cite{hirsch2000aurora}. The performance of the algorithms on the Aurora-2 database is a good indicator of how they might perform in the real world because Aurora-2's speech samples are mixed with noise collected from realistic settings such as streets, airports, and cars. Table \ref{tab:eval_pre_vad} is a list of existing SOTA VAD algorithms' accuracy scores on the testing set of Aurora-2. Unfortunately, as we can see, the highest-performing one is rVAD, which only achieves an accuracy score of 66.23\%, which is far from being usable in the real world. In other words, there exists solutions in the literature that do not work on datasets with deamplification, reverberation, and background noise. It is good to discover that these solutions are not likely to work at post-deployment time, because this helps us filter out existing solutions so that we won't use those solutions.

\begin{table}[]
\centering
\begin{tabular}{cc}
\toprule
Existing SOTA VAD         & Accuracy  \\
\hline
VQVAD       & 45.66\%          \\
Sohn et al. & 31.21\%          \\
Kaldi Energy VAD    & 11.72\%       \\
DSR AFE     & 40.07\%           \\
rVAD        & 66.23\%           \\
\bottomrule
\\
\end{tabular}
\caption{Evaluation of existing VAD algorithms on the Aurora-2 database. This Table is reported by Tan et al. \cite{tan2020rvad}.}
\label{tab:eval_pre_vad}
\end{table}

However, there was another algorithm, the Google Speech Recognition (GSR) algorithm, that had not been evaluated on a dataset that contained the three environmental distortions: reverberation, deamplification, and background noise. As a result, we next evaluated the Google Speech Recognition solution. we aimed to evaluate it in a comprehensive way to demonstrate that it would be robust against environmental distortions such as reverberation, deamplification, and background noise. Again, this is to set up the necessary condition to prove our hypothesis that a solution, during the pre-deployment stage, must be able to deal with the unique challenges given the real, designated environment in which it will be deployed. 

To do so, we collected a dataset that contains diverse environmental distortions: first, we collected the clean samples - samples that are not environmentally distorted, by having an individual talk next to the microphone for 5 minutes. The 5-minute clip was then sliced into 60 5-second segments, each of which is individually labeled as positive if it contained a human voice, or negative if it did not contain human voices. Note that the individual took long pauses intentionally to make sure that there were negative samples. Second, we collected the audio clips that were deamplified and contained background noise. To do so, we took copies of the clean samples. For each of the copies, we randomly deamplify them by m decibels (0 < m < 12) as per the practice of a previous work on emotion detection \cite{gao2021emotion}. Then, we randomly picked household sounds from the household ambience dataset \cite{mesaros2016tut}. Table \ref{tab:homenoise} lists the events that occur in the dataset. Note that each of the ambience sounds is greater than 5 seconds, so we randomly picked a segment from it that was 5-seconds long, and overlaid it with a deamplified clip. We repeated this process for all 60 deamplified clips. Third, we created the data for reverberated speech. To do so, we took another set of copies of the clean samples, and overlaid each of them with reverberation that was described by the combination of the three parameters: the wet/dry ratio $r$, diffusion $d$, and decay factor $f$. Finally, we created samples that are deamplified, noise-contaminated, and reverberated. To do so, we took a set of copies of the 60 deamplified and noise-contaminated samples, and overlaid them with the same reverberation effect as the samples that only contained reverberation effect and nothing more. In the end, we have 60 clean samples, 60 deamplified and noise-contaminated samples, 60 reverberated samples, and 60 samples that had all three environmental distortions. As a result, we claim that we created a dataset that was comprehensive enough to account for all three kinds of environmental distortions.

\begin{table}
    \centering
    
    \begin{tabular}{cc}
    \toprule
    Event & Instances\\
    \midrule
    (object) rustling & 60 \\
    (object) snapping  & 57 \\
    cupboard & 40 \\
    cutlery & 76\\
    dishes & 151\\
    drawers & 51\\
    glass jingling & 36 \\
    object impact & 250 \\
    people walking & 54 \\
    washing dishes & 84 \\
    water tap running & 47 \\
    \bottomrule
\end{tabular}
\caption{Events that are present in the background noise collected from real homes from the dataset \cite{mesaros2016tut}. All of them are covered in the process of contaminating audio samples with background noise. Note that this list do not include sounds from the tv, which are very important to make sure the robustness of the emotion detection model and conflict detection model.}
\label{tab:homenoise}
\end{table}

We evaluated GSR on the dataset that we just created. \textbf{GSR achieved an accuracy score of 95.83\%}, correctly classifying 230 out of the 240 samples each of which accounted for the environmental distortions to a certain degree. The high performance led us to decide to deploy GSR as our VAD model since, during the pre-deployment stage assessment, it is shown to be robust against the challenges that it is about to encounter in the real, designated environment: reverberation, background noise, and deamplification. 



\subsubsection{Post-deployment Stage Assessment}

Using post-deployment data on six completed dyads, we validate how well the chosen solution worked in practice. Table \ref{tab:eval_post_vad} shows the evaluation results of the VAD model on the dyads. We randomly select samples generated by each dyad during their deployment, and have human labelers label them if they are of human speech or not. We obtained 100 samples for all the dyads. The high performance of the VAD model indicates that this part of our system is highly effective at filtering out non-human speech samples such as background music (without lyrics) and footsteps. It is noted that the VAD does not filter out TV sounds if there is human speech in the sounds, such the voices of actors or news anchors. These unwanted sounds are filtered by the next model, SID.

The VAD model achieves an accuracy score of 94.0\% to 100\% on the six dyads. The high performance on post-deployment data validates our choice of the Google Speech Recognition in the pre-deployment phase. This implies that this VAD algorithm was originally made very robust to real world complexities. The high performance also indicates that, in order for the deployment to be successful, smart health groups using audio should perform pre-deployment tests with comprehensive real-world distortions. In addition, the high performance suggests that our hypothesis holds true - recall that our hypothesis is that, during the pre-deployment stage assessment, an about-to-be-deployed algorithm must be proven to overcome the challenges that are perceived to be present in the real, designated environment in order for it to perform well in said environment. The high performance on post-deployment data also indicates that it is sometimes possible to perform comprehensive and realistic pre-deployment testing to improve VAD post-deployment success.

\begin{table}[]
\centering
\scalebox{0.9}{
\begin{tabular}{ccccccc}
\toprule
          & Dyad 1  & Dyad 2  & Dyad 3 & Dyad 4 & Dyad 5 & Dyad 6\\
\hline
GSR & 100\%  & 100\%   & 94.0\%    & 95.0\%  & 100\%   & 98.0\% \\
\bottomrule
\\
\end{tabular}}
\caption{The evaluation results for the voice activity detection model on the dyads. The high accuracy scores achieved from the dyads indicate that the VAD algorithm (Google Speech Recognition) is highly effective at differentiating non-speech from human speech audio samples.}
\label{tab:eval_post_vad}
\end{table}

\subsection{Speaker Identification (SID)}

The SID model determines the identity of a speaker. The SID is a crucial part of the Acoustic System because we only want the voices of the caregiver and patient to be sent to the emotion and conflict detection models. However, in real deployments, voices from the TV and visitors must be filtered out. In this Section, we test SID model to answer this question: is it possible to perform comprehensive and realistic pre-deployment testing to improve post-deployment success? Note that the acoustical realisms that the SID faces, in addition to (non-speech) background noise, reverberation, and deamplification, also include sounds from the tv such as the dialogues from tv characters, for the presence of another person's voice in an audio sample could confuse the speaker identification model.

\subsubsection{Pre-deployment Stage Assessment}
During pre-deployment time, we investigated a state-of-the-art SID algorithm, the Google Speaker Identification API. However, the API asks us to input the maximum number of speakers there can be in a clip. This is impractical because a dyad can have the TV on and there could be many people's voices from the TV, or there may be multiple visitors. It is good to discover that this solution is not likely to work at post-deployment time, because this helps us filter out existing solution(s).

Now we describe how we test to make sure the about-to-be-deployed SID algorithm developed by Microsoft \cite{chen2022wavlm} is robust to environmental distortions such as reverberation, background noise, and deamplification. Again, this is to verify our hypothesis that for an algorithm to be successful in the real, designated environment, it must be able to overcome the challenges present in the real, designated environment during the pre-deployment stage. In our case, the challenges are reverberation, deamplification, non-speech background noise and TV sounds. Specifically, we have two persons, P1 and P2, each of whom spoke next to the microphone for 2.5 minutes. Then, for each of their voice files, we sliced it into 28 audio samples. Because these 56 (28$\times$2) samples were collected when the speakers were right next to the microphone, they were considered clean speech, free of the three types of environmental distortions. We needed to craft environmentally distorted samples out of the 56 clean samples to ensure that the testing samples accounted for both clean and environmentally distorted samples. To do so, we copy each of the 56 clips and deamplify them by randomly choosing a real number between 0 and 12 decibels. Then, we randomly chose a noise clip from Table \ref{tab:homenoise} as well as TV sounds we recorded using a microphone, out of which we randomly chose a consecutive 5-second segment to be overlaid with one of the copies. This guaranteed samples that were deamplified and contaminated with noise, and the last step was to reverberate it. Again, the reverberation effect is described by the three parameters: the wet/dry ration $r$, diffusion $d$, and decay factor $f$, as per the practice of a previous work \cite{salekin2017distant}. When we reverberated a (noise-contaminated and deamplified) copy, the values of $r$, $d$, and $f$ are randomly chosen. In total, we had 112 samples, 56 of which belonged to P1 and the other 56 belonged to P2. We fed the 112 samples to our SID model. \textbf{The SID model achieves an f1 score of 85.7\% on P1 and 92.8\% on P2}. The high performance of the SID model led us to believe that it was reasonably robust to reverberation, deamplification, background noise, and TV sound.


\subsubsection{Post-deployment Stage Assessment}

To validate post-deployment success, from all audio samples that our speaker identification algorithm identifies to contain the voice of the caregiver or the patient, or both, we randomly chose 28 from the first dyad, 28 from the second dyad, and 28 from the third dyad, 100 from the fourth dyad, 80 from the fifth dyad, and 100 from the sixth dyad. The results are reported in Table \ref{tab:eval_sid}. In the following sentences we describe how we obtain the f1 scores in Table \ref{tab:eval_sid}. For a sample, if it only contains the voice of the caregiver, then it is labeled as belonging to the caregiver; it if only contains the voice of the patient, then it is labeled as belonging to the patient. If it contains voices from both the caregiver and patient, then it is labeled as belonging to both. Otherwise, it labeled as belonging to neither. With this labeling scheme, we obtain the positives and negatives of the caregiver's voice and the positives and negatives of the patient's voice. The SID model can label a sample as belonging to the caregiver, belonging to the patient, or neither. As a result, we obtain the results in Table \ref{tab:eval_sid} in which we report the f1 scores to measure the performance of our SID model for both the caregiver and patient of each dyad. The SID model achieves an f1 score in the range of 93.1\% to 97.4\% for the caregivers and 91.6\% to 98.3\% on the patients in the six dyads. The high performance in Table \ref{tab:eval_sid} indicates that our SID algorithm is effective at picking out the voices by the caregiver and the patient in each home in their real home environment. Given that the SID algorithm was specifically assessed to see if it could overcome the challenges (reverberation, deamplification, and background noise) present in the real, designated environment (homes), we have shown that for an algorithm to be successful in the real, designated environment, it must be able to overcome the challenges present in the real, designated environment during the pre-deployment stage. The high performance of the SID during the post-deployment time suggests that our way to perform comprehensive and realistic pre-deployment is effective at improving post-deployment SID success. Note that we only validate the SID solution on the voices of the caregiver and patient of each dyad, because at post-deployment time, the SID solution filtered out voice samples that belonged to neither. As a result, we only have samples that are labelled by the SID solution as either the caregiver or the patient. For samples that made through the SID solution, we have the performance reported in Table \ref{tab:eval_sid}.

In Table \ref{tab:eval_sid_other} we report the f1 score of a model \cite{lecun1995convolutional} that we did not use because at pre-deployment time it achieves bad performance (an f1 score of 79.3\% on P1 and an f1 score of 71.2\% on P2). As we can see, this model also achieves bad performance on the post-deployment data. This indicates that at pre-deployment time, the model that performs badly also performs badly at post-deployment time.

\begin{table}[]
\centering
\scalebox{0.9}{
\begin{tabular}{ccccccc}
\toprule
          & Dyad 1  & Dyad 2  & Dyad 3 & Dyad 4 & Dyad 5 & Dyad 6\\
\hline
Caregiver & 94.5\%  & 97.4\%    &  95.7\%   & 93.1\%    & 92.0\%    & 89.2\% \\
Patient &   94.6\%  & 95.9\%    &  96.0\%   & 94.6\%    & 91.6\%    & 98.3\% \\
\bottomrule
\\
\end{tabular}}
\caption{The evaluation results for the speaker identification model on the dyads. The results are the f1 scores.}
\label{tab:eval_sid}
\end{table}

\begin{table}[]
\centering
\scalebox{0.9}{
\begin{tabular}{ccccccc}
\toprule
          & Dyad 1  & Dyad 2  & Dyad 3 & Dyad 4 & Dyad 5 & Dyad 6\\
\hline
Caregiver &   57.1\% & 71.4\%    & 83.6\%   & 52.6\%    & 44.3\%    & -\\
Patient &   74.8\%  & 75.9\%    &  79.5\%   & 77.7\%    & 74.2\%    & 97.0\%\\
\bottomrule
\\
\end{tabular}}
\caption{The post-deployment evaluation results for a speaker identification model that \textbf{we did not use} because it achieved bad performance pre-deployment time. As we can observe, its performance on all dyads is bad at post-deployment time.}
\label{tab:eval_sid_other}
\end{table}
\subsection{The Emotion Detection Models}

The emotion detection model detects the emotion of the speaker in a given audio clip. During the pre-deployment study, we looked into the state-of-the-art solutions for emotion detection. Table \ref{tab:sota_emotion_detection} shows the performance of several existing state-of-the-art solutions using various datasets of emotional utterances. In this Section, we aim to test the emotion detection model: is it possible to perform comprehensive and realistic pre-deployment testing to improve post-deployment success?

\begin{table}
  \begin{tabular}{ccccc}
    \toprule
    Work & Dataset(s)  & Accuracy\\
    \midrule

    Beard et al. \cite{beard2018multi} & SAVEE, CREMA-D &  41.2\% \\
    VGG \cite{simonyan2014very} & EMO-DB & 43.0\% \\
    Huang et al. \cite{huang2018stochastic} & RAVDESS, SAVEE &  60.8\% \\
    Ghaleb et al. \cite{ghaleb2019multimodal} & RAVDESS & 67.7\%  \\
    Ghaleb et al. \cite{ghaleb2019multimodal} & CREMA-D & 74.0\% \\
    SpeechBrain \cite{ravanelli2021speechbrain} & IEMOCAP & 78.7\%\\ 
    
  \bottomrule
\end{tabular}
\caption{State-of-the-art emotion detection algorithms on different emotional speech datasets. This Table is adapted from a table in the work \cite{gao2021emotion}. As we can see, the best performing state-of-the-art algorithm's accuracy is capped at 80\%. Since SpeechBrain achieves the highest performance, we select it as our emotion detection model.}
\label{tab:sota_emotion_detection}
\end{table}

\subsubsection{Pre-deployment Stage Assessment}
At first glance, Table \ref{tab:sota_emotion_detection} suggests that 4 solutions are not viable, but that SpeechBrain \cite{ravanelli2021speechbrain} is the best candidate among all the state-of-the-art algorithms, given its high performance on the dataset IEMOCAP \cite{busso2008iemocap}. But is it capable of overcoming the challenges (reverberation, deamplification, and background noise) that are present in the real, designated environment (a dyad's home)? To answer that question, we need to look into the dataset IEMOCAP \cite{busso2008iemocap}, on which it is evaluated. If the dataset has accounted for the challenges, i.e. during the data collection and processing process, the audio samples are touched by the effects of the three challenges, then we conclude that the evaluation result yielded by SpeechBrain indicates that it had overcome the three challenges perceived to be present in the real, designated environment that is a dyad's home. The dataset IEMOCAP \cite{busso2008iemocap} indicates that the audio clips are collected when there are furniture items in the room, instead of an empty acoustic studio, which suggests that the audio clips in IEMOCAP \cite{busso2008iemocap} are touched by the effect of reverberation. The audio clips in IEMOCAP \cite{busso2008iemocap} are \emph{not} collected where a speaker is right next to the microphone. This suggests that the audio clips in IEMOCAP \cite{busso2008iemocap} are touched by the effect of deamplification. Last but not least, IEMOCAP is not collected in a studio environment and there was no indication that the indoor background noise events such as footsteps were deliberately removed. Therefore, this suggests that the audio clips in IEMOCAP are touched by the effect of background noise. Consequently, we conclude that IEMOCAP's audio samples on which SpeechBrain is evaluated on account for the challenges that are perceived to be present in the real, designated environment in which the emotion detection algorithm (SpeechBrain) is to be deployed. Here, we set the stage to prove (once again) the hypothesis that for an algorithm to work well in the real, designated environment in which it is envisioned to be deployed, during pre-deployment stage, it must show that it is capable of handling the challenges that are perceived to arise in the real, designated environment. In the meantime, we have observed that algorithms such as Huang et al. \cite{huang2018stochastic} are not likely to work sufficiently at post-deployment time. It is good to discover that these solutions are not likely to work at post-deployment time, because this helps us filter out such existing solutions. 

Note that none of the state-of-the-art approaches in Table \ref{tab:sota_emotion_detection} include TV sounds as one of the acoustical realisms that they need to address. In the future, we plan to develop an emotion detection algorithm that takes TV sounds into consideration.

\subsubsection{Post-deployment Stage Assessment}

In the following paragraphs we describe how we evaluate our emotion detection detection model post-deployment. Out of the audio clips we collected from each dyad, we first select all samples that are classified by the emotion detection model and conflict detection model as anger speech. Then, we randomly select the same number of audio clips from all the samples by that dyad that are not classified as anger speech. Each of the audio clips is \textbf{manually labeled} based on the emotion in the clip by the labelers. Table \ref{tab:eval_emotion} describes our emotion detection model's performance on the labeled samples: For the 1st dyad, there are 233 samples. For the 2nd dyads, there are 392 samples. For the 3rd dyads, there are 281 samples. For the 4th and 6th dyads, there are 100 samples each. For the 5th dyads, there are 80 samples.

During the post-deployment stage assessment, researchers should let third-party labellers label the data to obtain ground truth, instead of letting the participants do the labelling, because the participants are not necessarily good at discerning their own emotion (and if they are in a verbal conflict) if they are not trained. The XYZ system detects if a person is in a verbal conflict or is angry. Initially, we sought to validate our system's performance by survey questions using EMA, similar to many other studies. However, we quickly found out that their responses did not always agree with the decision of the system. Who was wrong: the caregiver or our machine learning solutions? There is existing literature \cite{goerlich2018multifaceted} stating that people not trained to recognize their emotions are often bad at recognizing their own emotions. To investigate, we employed 5 labellers who are approved by the IRB to listen to and label the saved clips of the participants' voices. Their labelling suggests that in some cases, the labellers annotation did not agree with the participants' self reported emotional states. This data supports the claim that people are often bad at recognizing their emotions. We were able to verify the claim only because we planed, during pre-deployment time, to save all the raw data during the full deployment time. We found that post deployment labeling is better than EMA surveys and also supports determination of ground truth which, in turn, provides a better accuracy assessment of the acoustic classifiers. Determining ground truth from deployment time data is very important and often not done in many studies. 

The emotion detection model achieves an f1 score of 85.3\% to 97.4\% on the six dyads. According to Table \ref{tab:eval_emotion}, the emotion detection algorithm (SpeechBrain)'s performances in all six homes are satisfactory, highly efficient at identifying the emotions in each clip in each of the six homes. The success of the emotion detection algorithm demonstrated by Table \ref{tab:eval_emotion} proves our hypothesis: for an algorithm to be able to work satisfactorily post-deployment in a real, designated environment, it must demonstrate that it is able to overcome the challenges perceived to arise in that environment during pre-deployment time. The high performance of the emotion detection model at post-deployment time suggests that our way to perform comprehensive and realistic pre-deployment testing is effective at improving post-deployment success. Table \ref{tab:eval_emotion} also includes the performance of VGG \cite{simonyan2014very} which yields bad results (an accuracy score of 43.0\%) at pre-deployment time. As we can see, the model that achieves bad performance at pre-deployment time also achieves bad performance (an average of an f1 score of 50.1\%).

\begin{table}[]
\centering
\scalebox{0.85}{
\begin{tabular}{ccccccc}
\toprule
          & Dyad 1  & Dyad 2  & Dyad 3  & Dyad 4 & Dyad 5 & Dyad 6\\
\hline
SpeechBrain       & 88.8\%  & 87.2\%   & 91.8\%    & 92.9\% & 85.3\%  & 97.4\%   \\
VGG         & 44.6\%    & 65.4\%     & 48.3 \%  & 45.3\% & 53.3\% & 61.5\%  \\
\bottomrule
\\
\end{tabular}
}
\caption{The post-deployment evaluation results for the emotion detection model (SpeechBrain) as well as the VGG model \cite{simonyan2014very}, \textbf{which we did not use} because at pre-deployment time it achieves bad performance with an accuracy score of 43.0\%. As we can see, it also achieves bad performance on the post-deployment data. The measurement in this Table is f1 score.}
\label{tab:eval_emotion}
\end{table}

\subsection{The Conflict Detection Model}

For conflict detection, currently, there is no available conflict detection algorithm that is acoustics-based. Therefore, we developed our own conflict detection algorithm. In this Section, we aim to test on the conflict detection model: is it possible to perform comprehensive and realistic pre-deployment testing to improve post-deployment success?

\subsubsection{Pre-deployment Stage Assessment}
Here we briefly describe how the new algorithm we developed is trained and why the training process makes it specifically account for the (three) challenges that arise in the real, designated environment in which the algorithm is going to be deployed. The training and testing samples are from 19 couples and each sample is labeled conflict if the content of the sample indicates that the couple are in a verbal conflict. It is labeled non-conflict if the couple are not in a verbal conflict. Since the samples are already collected from home-environments, de-amplification and reverberation are accounted for, but the samples are free of background noise. As a result, we mix each of the samples with background noise by randomly selecting a segment from a randomly chosen indoor background noise sample in Table \ref{tab:homenoise} and overlaying that segment with each sample. Out of the samples, there are 3,072 in the training set and 1,009 in the testing set. As a result, both the training and the testing set accounts for a variety range of indoor environmental distortions. Since the training samples are touched by deamplification, reverberation, and background noise, our conflict detection algorithm trained on them is designed to be able to handle the three challenges (deamplification, reverberation, and background noise).

The conflict detection model's performance on the testing set achieves an f1 score of 93.1\%. The high performance suggests that the conflict detection is robust against environmental distortions such as reverberation, background noise, and deamplification. This help us set the stage to prove our hypothesis that, for an algorithm to work sufficiently in the real, designated environment in which challenges are perceived to be present, the algorithm must show that, during pre-deployment stage, it is able to handle the challenges. Our conflict detection algorithm has indicated that during pre-deployment stage, it is able to handle the three challenges: reverberation, deamplification, and background noise. Note that we do not include TV sounds as one of the acoustic realisms that the conflict detection model needs to address. In the future, we plan to develop a conflict detection algorithm that takes TV sounds into consideration.


\begin{table}[]
\centering
\begin{tabular}{ccccccc}
\toprule
          & Dyad 1  & Dyad 2  & Dyad 3 & Dyad 4 & Dyad 5 & Dyad 6\\
\hline
Ours       & 63.4\%    & 65.9\%    & 70.7\%  & 86.2\% & 82.7\% & 90.1\%\\
\bottomrule
\\
\end{tabular}
\caption{The evaluation results for the conflict detection model. The measurement is f1 score.}
\label{tab:eval_conflict}
\end{table}

\subsubsection{Post-deployment Stage Assessment}
In the post-deployment time, we seek to prove our hypothesis that, for an algorithm to work well post-deployment time in the real, designated environment in which it is going to be deployed, during pre-deployment stage it must show that it is capable of overcoming the challenges that are present in the real, designated environment. Our conflict detection algorithm has showed that it is capable of overcoming the challenges (it achieves an f1 score of 93.1\% pre-deployment time). However, is it going to work well in the post-deployment time?

Table \ref{tab:eval_conflict} shows our conflict detection algorithm's performance during the post-deployment time at the six homes. Now we explain how we obtain the f1 score results in Table \ref{tab:eval_conflict}. If a clip is labeled by the labelers such that it contains verbal conflict and the classifier also thinks this clip contains verbal conflict, then it is a hit. If the clip is labeled by the labelers as not containing verbal conflict and the classifier also thinks that it does not contain verbal conflict, then it is a hit. All other cases are misses (for example, the labelers think that a sample contain verbal conflict but the classifier fails to classify it as so). By looping through all samples produced by a dyad, we produce an f1 score on that dyad. From Table \ref{tab:eval_conflict}, we observe that the sixth dyad achieves the best performance with an f1 score of 90.1\% while the first dyad achieves the lowest performance with an f1 score of 63.4\%. For each of the dyads, we observe a drop in performance compared to 93.1\% obtained when the same model is evaluated on the dataset containing speech samples from the 19 couples. This indicates that despite our effort in mitigating environmental distortions, the effects of the environmental distortions such as room reverberation, background noise, and the deamplification effect are not fully mitigated. But the relatively satisfactory performance of the conflict detection model on dyads 4, 5 and 6 indicates that our way to perform comprehensive and realistic pre-deployment testing to improve post-deployment success is effective for exapected conditions.

We also investigate why the performance of the conflict detection model is lower in dyads 1-3 (f1 score of 63.4\% to 70.7\%). Upon communicating with the dyads, we learned that dyad 1 moved the system (which included the microphone) to the hallway which is very far away from the usual places that the participants were speaking. Dyad 2 had a construction team rennovating their home, so there was a lot of construction noise to confuse the conflict detection model. When we developed the conflict detection model, we did not take construction noises into consideration. In dyad 3, the caregiver's voice was always very low, almost inaudible, and our conflict detection model was not designed to handle such low-to-inaudible voice samples.

\subsection{Summary}

In this Section we briefly summarize our findings in Section \ref{sec:evaluation}. 

The first finding is that, to ensure post-deployment success of an algorithm, during pre-deployment time it must be rigorously tested on samples that are touched by the challenges that are perceived to be present in the real, designated environment during post-deployment time. This finding is confirmed by the pre-deployment stage assessment results and post-deployment stage assessment results of the VAD model, the SID model, the emotion detection model, and the conflict detection model. 

The second finding is that in the acoustic processing pipeline we should always go for off-the-shelf solutions first before developing your own algorithm. Many acoustic functions have been under study for many years and excellent solutions exist. Yet, rigorous testing is still required since not all of the available solutions will work for the environment where the system will be deployed. In our case, the VAD, SID, and emotion detection models are off-the-shelf. Since there are no state-of-the-art conflict detection models that \emph{only use (the prosody of the) voice} to detect verbal conflict, this has to be developed and will not have the luxury of having solutions refined many many researchers over many years. This likely is why the performance is lower than the well developed acoustic functions. 

\section{\emph{Adaptable} Add-on Components}

\label{sec:adaptability}
Deploying research systems over long time periods requires significant adaptability in many ways. Because the user is central, in this section we concentrate on adaptability in regards to the user interface.

\subsection{EMA Wording}
\label{sec:ema_wording}
EMA apps are common interfaces to humans. They are used to send messages to the participants and receive responses. In our system messages are the interventions or recommendations based on the output of the emotion, conflict detection, and reinforcement learning recommendation modules. 

During the pre-deployment time, we collaborated with a team of psychologists and specialists from the nursing field and determined four main categories of interventions/recommendations: the breathing exercise, which encourages the participants to take deep breaths to calm down, the timeout exercise, which encourages the participants to take a timeout/break from engaging with their loved one (the dementia patient), the mindfulness exercise, which encourages them to practice mindfulness, and enjoyable activities, which encourage them to partake activities that they enjoy. Each of these categories contains multiple subcategories. Our implementation  allows the exact wordings of the messages to be easily changeable in case during the deployment time the participants preferred different wording. 

Indeed, during the deployment time, we received feedback from the some participants that certain messages seemed insensitive and harsh, and the insensitivity and harshness of the messages actually discouraged them from implementing the recommended interventions. To fix this problem for these users, we went back to the EMA app and easily changed the wording based on the participants' complaints. After we made the changes to the wording for all participants, they no longer felt the wording as insensitive and harsh, and were encouraged by the messages to implement the interventions. 


\subsection{Personalized Recommendations}
We have previously stated in Section \ref{sec:ema_wording} that there are four categories of interventions or recommendations. In this Section, we describe how we make the personalized recommendations adaptable during pre-deployment time and verify if our strategy (that makes the recommendations adaptable) succeeds during post-deployment time.

We have also stated that during pre-deployment time we made the EMA app to be easily adjustable in case we need to change the interventions or recommendations. The EMA app is easily adjustable because it reads text questions (such as the list of enjoyable activities for the participants) from a database and to change that list, we just need to go to the database instead of recompiling the app every time we make changes to the list of text surveys and interventions. The adaptability of the EMA app and, therefore, the XYZ-W system came in handy when we discovered that participants want more specific recommendations, especially under the category of enjoyable activities. In other words, instead of a generic ``now it's time to do some enjoyable activities,'' they wanted the enjoyable activities to be more specific and personalized such as ``now it's time to play with the family cat.'' To accommodate this, we asked each participant to provide us a list of their personalized enjoyable activities. Because the EMA app was made to be easily adjustable, the integration of the personalized enjoyable activities for each participant was quick and easy. 

The participants of all six dyads reported that they liked the personalization (we asked them through surveys during our interviews with them). For smart health apps that make recommendations, we feel that it is imperative to be able to change them to accommodate the personal needs of the participants.


\subsection{Positive Feedback}
Before we deployed our XYZ-W System, we decided to only send out at the end of the day post-recommendation surveys that asked the participants if they had implemented the recommendations. Then, we received requests from the participants that they would like some positive feedback after they implemented the recommendations. The positive feedback should acknowledge the effort they put in to adhering to the recommendations and remind them of the importance of implementing the recommendations (to improve their mental health and lessen their care-giving burden). Again, because the EMA app was designed at pre-deployment time to be easily changed, integrating the feature of positive feedback into the existing system was quick and easy. The participants of all six dyads reported to have liked the positive feedback.  


\subsection{Summary}

In this Section we briefly summarize our findings in Section \ref{sec:adaptability}. The main takeaway message in Section \ref{sec:adaptability} is that researchers need to anticipate the need of the participants and make sure that their technology (in our case, the add-on components to the XYZ System), is adaptable to those needs. For example, we initially set the positive feedback times to be in the morning and in the evening, but participants report that they want the positive feedback more often. Since we have anticipated that they might have the need (to want to see the positive feedback more often), we have designed our add-on components in such a way that allows the change to be easily made. 
\section{Privacy Concerns}
Developing new research systems that will be deployed in homes require solutions for privacy issues during recruitment and during deployment. This is especially true when using sensing that can be considered invasive such a microphones. Our XYZ system is privacy-conserving, meaning that it concentrates on allowing users to benefit from its functionalities without releasing their voice data in its original form.

\subsection{Pre-deployment Stage Preparation}

During pre-deployment a set of privacy mechanisms were implemented. 
This included that users could set the day start and end times when the system is operational, turn off the system at any point, e.g., when a visitor was there, all voice from non-registered speakers would not be recorded, and that the content of speech would not be recorded, i.e.,  only prosidy features would be recorded.

During recruitment, the above privacy mechanisms were fully described to potential dyads. Also, using a graph and tables we showed them what prosidy means.
Nevertheless, one in four care-giving dyads who rejected participation in the study worried about the privacy issue despite the thorough explanation of the study procedures. Basically, once we explained that our system listens to them for anger/conflict events, it caused some of the prospective dyads to think about privacy invasion and they declined to participate in the study.

Overall, Table \ref{tab:human_behavior_privacy_recruitment} describes the reasons that potential participants did not choose to participate in our study: 3 potential participants did not meet the criteria to be included in our study. 15 potential participants lost interest in participating and did not reply to our call for screening that they must pass in order to be included in our study. 12 potential participants decided upfront with us during our communication with them that they were not interested. 10 potential participants found the study to likely be burdensome. 2 potential participants in our follow-up calls with them indicated that they are no longer interested in the study. 9 potential participants were no longer interested to participate due to various issues such as the patient passing away. 17 potential participants refused to participate out of the concern for privacy issues. Privacy is the main reason for 25\% of those interviewed (the largest percentage among all categories of the reasons to not participate). It can be expected that when deploying sensing systems in homes some people will reject the system. With enough privacy mechanisms we were able to limit those not willing to participate due to privacy to 25\%.

\begin{table}[]
\centering
\begin{tabular}{cc}
\toprule
  Reason for Rejection & Number(\%) \\
\hline
Did not meet inclusion criteria &	3 (4.41\%) \\
Didn't reply to call for screening &	15 (22.06\%) \\
Not interested &	12 (17.65\%) \\
Burdensome &	10 (14.71\%) \\
No longer wanted to participate &	2 (2.94\%) \\
Other &	9 (13.23\%) \\
\textbf{Worried about privacy issue} &	\textbf{17 (25.00\%)} \\
\hline
Total &	68 (100\%) \\

\bottomrule
\\
\end{tabular}
\caption{The reasons that potential participants did not join our study. 25\% of the potential participants refused to participate in our study because they were worried about privacy since our system would be listening to their conversation.}
\label{tab:human_behavior_privacy_recruitment}
\end{table}


\subsection{Post-deployment Stage Assessment}



Using a survey after the 4 month study, five out of six of the recruited dyads agreed that their privacy concerns are addressed and they have had no issue regarding privacy. However, one dyad still has some concerns over their privacy. In this home, the Alzheimers' patient (not the caregiver) pulled out the wires that connect our equipment to the power source, for he feared that during the time the system is set to not monitor them, the system is still actively listening. 
We countered this distrust our system by showing these participants the data we collected: namely, we showed them the timestamps of the voice samples that we collected so they knew that outside of the active hours of the system that we were true to our word and the system did not listen to them. We also showed them amplitude plots of the data we collected where there are no words. We also showed them that there did not exist files which recorded the transcriptions of their conversation, since some concerns had been raised such that the participants were worried about being transcribed.

\subsection{Summary}
Privacy concerns are likely to arise when researchers deploy their technology in the home environments. During the pre-deployment stage, researchers must implement privacy mechanisms even if they are not the core purpose of the system. These mechanisms must be explained to participants. We believe that the mechanisms we describe can be used in many systems that utilize speech of the participants. 
Finally, the responses from the dyads at post-deployment time provide some evidence that privacy can be addressed to users' satisfaction. 
\section{Conclusion}
As research projects (not pilot studies), smart health groups often design novel smart technologies to be deployed in people's homes. However, the integration of novel smart technology from the less complex lab environments to the more complex real environments poses many challenges. In this paper, using the XYZ system as a case study, we present and evaluate various techniques for acoustic pipelines, adaptability of the user interface, and privacy that increase deployment time success. 

\bibliographystyle{ACM-Reference-Format}
\bibliography{sample-base}


\end{document}